\documentstyle[epsfig,twocolumn,aps,prl,epsf,floats]{revtex}
\begin{document}
\draft
\flushbottom
\twocolumn[
\hsize\textwidth\columnwidth\hsize\csname @twocolumnfalse\endcsname
\title{Vortex Structure in Underdoped Cuprates}
\author{Patrick A. Lee and Xiao-Gang Wen}
\address{Center for Materials Science and Engineering
and Department of Physics, MIT, \\Cambridge, MA 02139}
\widetext
\date{ Aug. 29,\ 2000}
\maketitle
\tightenlines
\widetext
\advance\leftskip by 57pt
\advance\rightskip by 57pt
\begin{abstract}
In underdoped cuprates the normal state is highly anomalous and is
characterized as a pseudogap phase.
The question of how to describe the ``normal'' core of a superconducting
vortex is an outstanding open
problem.  We show that the SU(2) formulation of the $t$-$J$ model provides
a description of the normal
state as well as the vortex core.  Interestingly, the pseudogap persists
inside the core. We also found that it is likely
that the core consists of a state which breaks
translational symmetry due to the existence of staggered current which
generate staggered magnetic field with very slow dynamics.  
This staggered flux state is
likely to be the ground state for magnetic fields higher than $H_{c2}$.
Experiments to test this picture are proposed.
\end{abstract}
\pacs{ }
]
\tightenlines
\narrowtext

\section{Introduction}
It is now widely appreciated that high $T_c$
superconductors are fundamentally different from
conventional superconductors in that they emerge by
introducing doped holes into a Mott insulator.  This
contrast is most apparent in the underdoped region
where the density $x$ of doped holes is small.
Experimentally, this is also the regime where the
physical properties are most anomalous.\cite{PAL99}  Much
attention has been focused on the normal state, which
is characterized by a pseudogap regime below a relatively high temperature
$T^\ast\approx 300K$.
The pseudogap appears in spin excitations and in tunneling and ARPES
experiments.  The
superconducting state is anomalous as well, in that the superfluid density is
proportional to the hole density $x$ and not the electron density (Fermi
surface area)
$1-x$ as in conventional superconductors.  Recently, it has become possible
to perform
STM tunneling in the superconducting state and probe the electronic
structure of the
vortex core.\cite{Renner,Pan}  This raises the following interesting
question.  Common sense would
indicate that the vortex core should be made up of the normal state and one
would expect
the pseudogap, {\it i.e.} a dip in the tunneling density of states, to persist
in the core
region.  This is in fact what is seen experimentally.  Yet a conventional
description of
a vortex core requires that the order parameter vanishes inside the core,
which is
usually accompanied by the vanishing of the energy gap.  Thus it is clear
that the
electronic structure of the vortex core in the underdoped region is
qualitatively
different from that given by conventional theory.  This point was made
eloquently in a
recent paper by Franz and Te\u{s}anovi\'{c} (FT).\cite{FT}

It is clear that any attempt to model the underdoped vortex core must
include the physics of the
proximity to the Mott insulator, {\it i.e.} the strong correlation physics. One
of the few
analytic tools available for this purpose is the slave boson method used to
treat the
constraint of no double occupation in a strong correlation model such as
the $t$-$J$
model.  FT employed the U(1) formulation of this theory, where the electron
operator
$c_{i\sigma}$ is written as $c_{i\sigma} = f_{i\sigma}b_i^\dagger$ and the
no double
occupation constraint is replaced by $f_{i\sigma}^\dagger f_{i\sigma}+
b_i^\dagger b_i = 1$, which
is in turn accomplished by the introduction of a U(1) gauge field
$\bbox{a}$.  In mean
field theory the pseudogap gap state is described by a pairing of the fermions,
$\Delta_f(\hat{\eta}) = \langle f_{\uparrow,i}f_{\downarrow,i + \hat{\eta}}
\rangle$,
where $\hat{\eta}$ is a nearest-neighbor vector and $\Delta (\hat{\eta})$
has $d$
symmetry.\cite{Kotliar,Suzumura}  The superconducting state is described by
bose condensation of the
bosons
$\langle b \rangle \neq 0$.  $\Delta_f$ is not gauge invariant and the
onset of the
pseudogap is merely a cross-over, but the appearance of $\langle b \rangle
= b_0 \neq 0$
triggers the appearance of the superconducting pairing amplitude
$\langle c_{\uparrow i}c_{\downarrow i + \hat{\eta}} \rangle = b_0^2 \Delta_f$
which is gauge invariant and physical.  Within
this theory, FT propose a description of the vortex state where the bosonic
amplitude
$\langle b \rangle$ vanishes inside the core but the fermion pairing
amplitude $|
\Delta_f |$ remains finite.  Since the electronic spectrum is given by the
fermion
dispersion, the core will retain the energy gap, just as in the pseudogap
state.

Upon
closer examination, FT pointed out that this solution requires that the
gauge field has
negligible restoring force, {\it i.e.} a ``Maxwell'' term of the form $\sigma
({\bf \nabla}
\times \bbox{a})^2$ must have very small coefficient $\sigma$.  This
requirement is in fact related to
a problem discussed by Sachdev\cite{Sachdev} and by Nagaosa and
Lee\cite{Nagaosa/Lee} some time ago.
Due to the existence of two fields $\Delta_f$ and $\langle  b  \rangle$, it
is possible to
construct several kinds of vortices.  The field $\Delta_f$ is minimally
coupled to $\bbox
{a}$ in the form $\left|\left( \frac{\nabla}{i} + 2 \bbox{a} \right)
\Delta_f \right|^2$,
whereas the field $\langle b \rangle$ is coupled to a combination of
$\bbox{a}$ and the
electromagnetic field $\bbox{A}$ in the form $\left| \left( \frac{{\bf
\nabla}}{i} + \bbox{a} -
\frac{e}{c}\bbox{A} \right) \langle b \rangle \right|^2$.

The different kinds of vortices are as follows:

\begin{itemize}
\item[(i.)] A vortex carrying the conventional $hc/2e$ flux quantum. A gauge
vortex carrying half a flux
quantum
$\frac{1}{2}h$ is generated so that $\langle b \rangle$ has no singularity.
The phase of $\Delta_f$
winds by $2 \pi$ and its amplitude vanishes in the core.  This is just like the
conventional vortex in that the energy gap vanishes inside the core.  This
describes the
optimal or overdoped region.

\item[(ii.)] An $hc/e$ vortex.  This involves no winding of
$\Delta_f$ and no gauge flux.  The advantage is that $| \Delta_f |$ is
finite in the
core and the pseudogap is preserved.  This state is energetically favorable
because the
cost of the boson vortex is small for small $x$.  The price one pays is
that because the
boson carries charge $e$, this vortex carries a double superconducting flux
quantum
$\frac{hc}{e}$.  This has so far not been observed.

\item[(iii.)] The FT vortex.  A third possibility proposed by FT
is that a flux tube for the gauge field $\bbox{a}$ carrying $-\frac{1}{2}h$
gauge flux is
attached to the core.  Now the $\bbox{A}$ flux can be a conventional  flux
quantum
$\frac{hc}{2e}$ and the phase of $\langle b \rangle$ winds by $2\pi$, with
$\langle b
\rangle = 0$ in the core.  On the other hand, $\Delta_f$ sees only a flux
tube and
remains non-zero in the core.  It is this latter requirement which
forces the $\bbox{a}$
flux to be a flux tube, {\it i.e.} confined to a lattice plaquette.  
\end{itemize}

Actually
this possibility was considered by Nagaosa
and Lee and dismissed because the energy cost of a flux tube is large in
the presence of a Maxwell term.
The point is that the theory for $\Delta_f$ and $\langle  b \rangle$ must
be considered as a low energy
effective action, and terms allowed by symmetry such as the Maxwell term
will be generated by eliminating
the high energy degrees of freedom.  We expect the energy of the flux tube
will be of the order of the cut-off
scale, {\it i.e.} the fermion band width $J$.
This will make this kind of vortex very
costly in energy compared with the $hc/e$ vortex in the limit of small $x$.

FT appealed to the papers by Nayak\cite{Nayak} and
D.H. Lee\cite{DHLee} to justify setting $\sigma = 0$.  
Even assuming for the sake of argument that $\sigma$ vanishes and the 
$\bbox{a}$ flux tube costs no energy,
the FT vortex still had a core energy at least of order $J$. Although the
pairing field $\Delta_f$ cannot see the $h/2$ flux tube of $\bbox{a}$, the
fermions see the flux tube. The mismatch of phase $\pi$ at the lattice
scale in the fermion wave function will cost an energy of order $J$.
(Actually, this is why the $h/2$ flux tube costs an energy of order $J$
as discussed above.)
In order to reduce this energy cost, $\Delta_f$ likes to vanish
in a region of size
of coherent length $\xi_F\sim v_f/\Delta_0$, where $\Delta_0$ is the spin gap. 
Such a vortex has a core very similar to the standard BCS vortex (the case (i)
mentioned above). The fermion
contribution to the core energy is reduced to a value of order $\Delta_0$.


We should add that recently Senthil and Fisher\cite{Senthil.a} proposed a
model of the vortex based on their
Z(2) gauge theory which carries $\frac{hc}{2e}$ flux quantum and contains a
pseudogap in the core.  This is
accomplished by attaching a Z(2) vortex to the core.  Senthil and
Fisher\cite{Senthil.b} recently showed how
the Z(2) gauge theory can be placed in the context of the U(1) theory and
it becomes clear that their
model of the vortex is
intimately related to that of FT.  Senthil and Fisher combine the phase of
the boson and half that of
$\Delta_f$ to form the phase of the ``chargon'' which bose condenses.  The
Z(2) vortex is then the
residue of the half flux tube of FT.  The Z(2) vortex is also localized to
a lattice plaquette and has an energy
gap which Senthil and Fisher identify with the pseudogap scale.  This also
renders this vortex costly in
comparison with the $hc/e$ vortex (where the chargon winds by $4\pi$) in
the limit of small $x$.  Thus we
conclude that models based on U(1) mean-field theory still have
difficulties coming up with a stable $hc/2e$
vortex with a pseudogap core in the limit of small doping.

Several years ago, we introduced an alternate
formulation of the constraint in the $t$-$J$ model called the SU(2)
theory.\cite{WenLee}  This model is
designed to connect smoothly to the Mott insulator at half-filling, in that
the SU(2) symmetry known to be
present at half-filling is preserved for finite doping.  The SU(2)
mean-field theory should have a better chance of
describing the small doping limit.  In this paper we show that this theory
leads naturally to a stable $hc/2e$
vortex in the underdoped limit.  The spin gap is finite both inside and
outside the vortex core. Possible experimental consequences are explored at
the end of the paper.  

\section{Review of the SU(2) Formulation}

First we summarize some of the salient features of the SU(2) formulation.
This is well understood in the
undoped case, where SU(2) doublets $\psi_{\uparrow j} = \left(  f_{\uparrow
j},f_{\downarrow j}^\dagger
\right)$   and $\psi_{\downarrow j} = \left(  f_{\downarrow j},
-f_{\uparrow j} ^\dagger \right)$  were
introduced on each site $j$ to represent the destruction of spin up and
spin down in the subspace of one
fermion per site.\cite{Affleck,Dagotto}  Wen and Lee extended the SU(2) 
formulation away from half filling by
introducing a doublet of bosons $h_j = (b_{1j},b_{2j})$.  The physical
electron is represented as an SU(2)
singlet formed out of the fermion and boson doublets: $c_{\sigma j} =
\frac{1}{\sqrt{2}} h_j^\dagger
\psi_{\sigma j}$.  The constraint of no double occupation is enforced by
projecting onto the SU(2)-singlet
subspace of the extended $h_i,\psi_{\sigma i}$ Hilbert space.  On each site
there are three such singlets,
corresponding to
$|{\rm spin \, up} \rangle = f_\uparrow ^\dagger |0\rangle$, $|{\rm spin \,
down} \rangle = f_\downarrow
^\dagger |0\rangle$ and
\begin{equation}
|{\rm hole} \rangle = \frac{1}{\sqrt{2}} \left(  b_1^\dagger + b_2 ^\dagger
f_\uparrow
^\dagger f_\downarrow^\dagger
\right) |0\rangle  .
\end{equation}
The role of the two bosons can be visualized as follows.  In contrast to
the U(1) formulation, the fermions
may remain at half-filling upon doping.  Then a typical fermion
configuration will contain spin-up or spin-down
singly occupied sites, as well as empty and doubly occupied sites.  The latter 
sites are both spin singlets and have the correct spin quantum number for a 
vacancy.  The $b_1$ boson is used to mark the empty site and
the $b_2$ boson the doubly occupied site, and both $b_1$ and $b_2$ carry
unit charge.  This picture is a
bit over simplified, in that it is a linear superposition given by Eq. (1)
which correctly specifies a physical hole.

In order to perform the projection to SU(2) singlet, three sets of gauge
fields $a_{0j}^\ell$, associated with
the three Pauli matrices $\tau^\ell$,$\ell = 1,2,3$, are needed.  These are
the generalization of the time
component of the gauge field $a_{0j}$ in the U(1) formulation.  The
exchange and hopping terms are
decoupled to give the mean-field Hamiltonian,

\begin{eqnarray}
H = \sum_{\langle jk \rangle}
&& \left(
J \psi_{\alpha j}^\dagger U_{jk}\psi_{\alpha k} + t h_j^\dagger U_{jk}h_k +
c.c.
\right) \nonumber \\
&& + \sum_j a_{0j}^\ell \left(  \frac{1}{2} \psi_{\alpha j}^\dagger
\tau^\ell \psi_{\alpha j} + h_j^\dagger
\tau^\ell h_j  \right) \nonumber \\
&& -\mu \sum_{j} h_j^\dagger h_j + \frac{J}{2} \sum_{<jk>}Tr
(U_{jk}^\dagger U_{jk} )
\end{eqnarray}
The matrix $U_{jk}$ is given by
\begin{equation}
U_{jk} =
\left(
\begin{array}{c}
-\chi_{jk}^\ast  \\ \Delta ^{f^\ast}_{jk} \end {array}
\begin{array}{c}
\Delta^f_{jk} \\ \chi_{jk}
\end{array}
\right)
\end{equation}
where
\begin{eqnarray}
\chi_{jk} &=& \langle   f_{\alpha j}^\dagger  f_{\alpha k} \rangle \nonumber \\
\Delta_{jk}^f &=& \langle  \epsilon_{\alpha\beta}f_{\alpha j}f_{\beta k}
\rangle
\end{eqnarray}
The hole density is $\langle  b_1^\dagger b_1 + b_2^\dagger b_2 \rangle =
x$ and is enforced by the
chemical potential $\mu$.  The Lagrangian associated with Eq. (2) is
invariant under the local SU(2) gauge
transformation
\begin{eqnarray}
\psi_{\alpha j}  \rightarrow  g_j^\dagger\psi_{\alpha j} \,\,\, , \,\,\,
h_j  & \rightarrow  & g_j^\dagger h_j
\,\,\, , \,\,\, U_{jk} \rightarrow g_j^\dagger U_{jk}g_k \,\,\, , \nonumber \\
a_{0j}^\ell \tau^\ell & \rightarrow & g_j^\dagger a_{0j}^\ell\tau^\ell g_j
- g_j\partial_\tau g_j^\ell
\end{eqnarray}
where $g_j = \exp(i\bbox{A}_j\cdot \bbox{\tau})$ is a space and $\tau$
dependent $2\times2$
matrix that represents an SU(2) group element.

In Ref. \cite{WenLee} the SU(2) mean-field theory was worked out by making
the approximation that
$a_{0j}^\ell$ is independent of space and $\tau$.  Of special interest is
the pseudogap phase
which occupies the low doping part of the phase diagram. 
(Note that the pseudogap phase is called 
staggered flux or $s$-flux phase in the $SU(2)$ theory of
Refs. \cite{WenLee} and \cite{Lee etal}. Despite its name, the $s$-flux phase
in the $SU(2)$ theory is translation invariant and has no staggered
physical magnetic field. In this paper, we will reserve
the name ``staggered flux phase" for the staggered flux phase in the
$U(1)$ theory, which does have staggered physical magnetic field.\cite{MA}
We use ``spin-gap phase'' to refer to what we previously called 
the staggered flux phase in the $SU(2)$ theory.)
 In the spin-gap phase 
$a_{0j}^\ell = 0$ with finite
nearest-neighbor $\chi_{jk} = \chi$ and
$\Delta_{jk}^f  = \pm \Delta^f$ for $\hat{x}$ or $\hat{y}$ bonds.  This
resembles the fermion pairing phase
in the $U(1)$ theory.  However, due to the gauge symmetry given by Eq. (5),
the same mean-field state can be
constructed with the choice
\begin{eqnarray}
U_{j,j+\hat{x}} &=& -i\chi - (-1)^{j_x+j_y}\Delta_f\tau_3 \nonumber \\
U_{j,j+\hat{y}} &=& -i\chi + (-1)^{j_x+j_y}\Delta_f\tau_3
\end{eqnarray}
This resembles the staggered flux phase in U(1) mean-field theory\cite{MA}
 because
the hopping matrix elements are
complex and the sum of the phase angle around a plaquette gives a flux
which alternates in sign, as shown in
Fig. 1.  Equation (6) and the fermion pairing state Eq. (3) both give the
fermion dispersion,
\begin{equation}
E_\pm = \pm \left( \varepsilon^2(\bbox{k}) + \eta^2  (\bbox{k}) \right)^{1/2}
\end{equation}
where
\begin{eqnarray}
\varepsilon(\bbox{k}) &=& -2J \chi \left ( \sin k_xa + \sin k_y a \right ) \\
\eta(\bbox{k}) &=& 2J\Delta_f \left( \sin k_x a - \sin k_y a \right)  .
\end{eqnarray}
Due to our gauge choice, this dispersion is shifted by $\left(
\frac{\pi}{2},\frac{\pi}{2} \right)$ compared
with the more conventional parameterization which has a maximum $\Delta_0 =
J\Delta_f$ at
$(0,\pi),(\pi,0)$ and nodes at $\left( \pm \frac{\pi}{2}, \pm \frac{\pi}{2}
\right)$.  The boson dispersion is
the same except that $J$ is replaced by $t$.

\begin{figure}
\epsfxsize=1.5truein
\centerline{
\epsffile{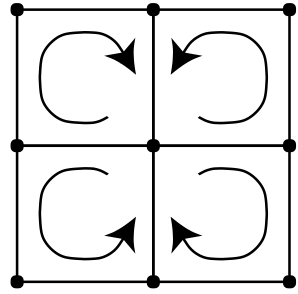}
}
\vspace{0.5cm}
\caption{Pictorial representation of the staggered flux state.  The hopping
integrals are complex in such a
way that the sum of the phase angle around a plaquette yields a net flux
which alternates in sign from
plaquette to plaquette.  This gives rise to circulating fermion currents on
the bonds as indicated by the
arrow.  In the presence of hole doping and condensation of the bosons,
circulating physical hole currents
appear.  We refer to this state as the staggered flux state. }
\end{figure}

The breaking of translation symmetry shown in Eq. (6) is a gauge artifact,
because Eq. (6) is mapped onto
Eq. (3) which is translationally invariant by a site dependent SU(2)
transformation.\cite{WenLee}  Thus in
the SU(2) mean-field theory, the spin-gap phase includes fluctuations
between the pairing state, the
staggered flux state, and many other states in the U(1) formulation.  Note
that the mean-field ansatz given in
Eq. (6) is itself invariant under a $\tau_3$ rotation.  Thus the SU(2)
symmetry has been broken down to U(1)
in the spin-gap phase.

\begin{figure}
\epsfxsize=2.0truein
\centerline{
\epsffile{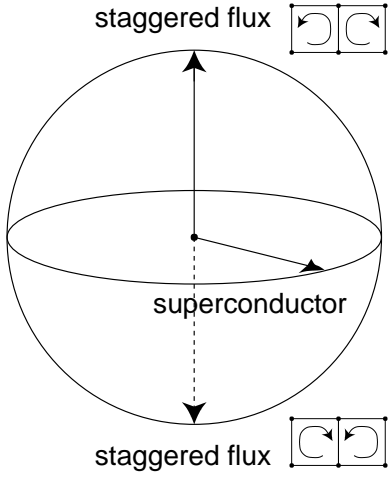}
}
\vspace{0.5cm}
\caption{The isospin quantization axis $\bbox{I}$ represents different
states depending on its orientation.  In
the north and south poles, it represents the staggered flux states.
These are two degenerate states with
the current pattern shifted by one lattice constant.  In the equator it
represents the $d$-wave
superconductor.  Vectors connected by rotation around the $\hat{z}$ axis
are gauge equivalent and
represent the same physical state. }
\end{figure}
In Ref. \cite{Lee etal} this point of view was clarified and the
approximation of constant
$\bbox{a}_0$ improved by introducing a nonlinear $\sigma$ model description
in terms of a
slowly varying boson field.  The idea is that at low temperatures the
bosons are nearly bose condensed to
the bottom of the boson bands and are slowly varying in space and time.  On
the other hand, the fermions
have a short coherence length $\xi_F = v_f/\Delta_0$ which is the lattice
scale because $\Delta_0 \sim J/3$.
Then the fermions follow the local boson field and can be integrated out,
after choosing an
${\bf a}_0$ field which minimizes the action locally.  The result is an
effective Lagrangian
which depends only on the local boson field.  It is convenient to choose
the fermion mean field in the
staggered flux representation given by Eq. (6), because the symmetry
breaking pattern from SU(2) to U(1)
is manifest.  The bottom of the boson band is at $\left( \frac{\pi}{2},
\frac{\pi}{2} \right)$ and we write
$h_j=\tilde{h}_j \exp \left(  -i(j_x + j_y)\frac{\pi}{2}  \right)$ and
consider $\tilde{h}_j$
 to be slowly varying.  At low temperatures,
$\tilde{h}_j^\dagger\tilde{h}_j = x$ and we write
\begin{equation}
\tilde{h}_j = \sqrt{x}
\left(
\begin{array}{c}
z_{j1}\\
z_{j2}
\end{array}
\right)
\end{equation}
where $\sum_\alpha \left| z_{j\alpha} \right|^2 = 1$ and are parametrized by
\begin{eqnarray}
z_{j1} &=& e^{i\alpha} e^ {-i\frac{\phi}{2}} \cos \frac{\theta}{2} \\
z_{j2} &=& e^{i\alpha} e^  {i\frac{\phi}{2}} \sin \frac{\theta}{2}
\end{eqnarray}
The phase $\alpha$ is the overall U(1) phase which couples to the
electromagnetic field.  The angles
$\theta$ and $\phi$ are best visualized by introducing the isospin
quantization axis $\bbox{I}$.
\begin{equation}
\bbox{I} = z_\alpha^\ast \bbox{\tau}_{\alpha\beta} z_\beta = \left(
\sin\theta \cos\phi, \sin\theta\sin\phi, \cos\theta
\right)
\end{equation}
{\it i.e.} $\theta$ and $\phi$ are the polar angles of the quantization axis.
The physical state depends on the
orientation of the vector $\bbox{I}$, as shown in Fig. 2.   $\bbox{I}$
pointing along $\hat{z}$ corresponds
to the staggered flux state in the U(1) formulation.  The polarization of
the boson field generates a nonzero
$a_0^3$, which corresponds to a shift in the chemical potential for the
fermions.  This in turn converts the
nodes at $\left(
\frac{\pi}{2}, \frac{\pi}{2} \right)$ into a small Fermi surface pocket.
This state breaks translational
symmetry and is characterized by a staggered pattern of {\it physical} hole
current distribution as shown in
Fig. 1.  We shall refer to this state as the staggered flux state.
$\bbox{I}$ pointing along
$-\hat{z}$ describes the same physical state except that the current
pattern in shifted by a unit cell.  On the
other hand,
$\bbox{I}$ in the
$x$-$y$ plane corresponds to a
$d$-wave superconductor state (with a finite chemical potential) which does
not break translational
symmetry.  This is not obvious in the representation given in Eq. (6) and
it is best seen by performing a space
dependent SU(2) rotation according to Eq. (5) to the representation given
in Eq. (3).  Note that $\bbox{I}$
pointing anywhere in the $x$-$y$ plane  is gauge equivalent and corresponds
to the same physical $d$-wave
superconducting state.  This is a consequence of the residual U(1) symmetry
of Eq. (6) and any vectors
$\bbox{I}$ related by a $\phi$ rotation represents states that are gauge
equivalent.  On the other hand,
$\bbox{I}$ pointing somewhere in between the north pole and the $x$-$y$
plane are distinct states with
both superconducting pairing and translational symmetry breaking.  This
state of affairs is summarized  by an
effective Lagrangian as derived in Ref.\cite{Lee etal}.   It takes the form
of an anisotropic $O(4)$
$\sigma$-model $(|z_1|^2 + |z_2|^2 = 1)$ coupled to gauge fields.
 For the purpose of this paper we restrict our attention
to time independent variation and the Lagrangian takes the simplified form
\begin{eqnarray}
L_{\rm eff} &=&  xt \left|D_j z\right|^2 + \frac{x^2J}{2} \left[
\frac{4}{c_1}|z_1z_2|^2  + \frac{1}{c_3} \left( |z_1|^2 - |z_2|^2\right)^2
\right]   \nonumber\\
&& + \frac{1}{2} a_j^{(3)} \Pi_{jk}a_k^{(3)}
\end{eqnarray}
where
\begin{equation}
D_j = \frac{\partial}{\partial r_j} + ia_j^{(3)}\tau^{(3)} - i\frac{e}{c}A_j
\end{equation}
is the covariant spatial derivative $(j = x,y)$,  $a_j^{(3)}$ is the
spatial component of the $\bbox{a}^{(3)}$
gauge field, and $c_1$ and $c_3$ are numerical constants of order 1.  
Since the SU(2)
symmetry has been broken
down to U(1) by Eq. (6), the
$\bbox{a}^{(2)}$ and
$\bbox{a}^{(3)}$ gauge fields are massive by the Higgs mechanism and have
been ignored. The first term in
Eq. (14) is the boson kinetic energy minimally coupled to the remaining
$\bbox{a}^{(3)}$ gauge field and the
electromagnetic field $\bbox{A}$.  The second term is a phenomenological
term introduced to describe the
difference in energy between the superconducting state and the staggered
flux state, so that the
quantization axis prefers to lie in the $x$-$y$ plane.  The third term
comes from integrating out the fermion
degrees of freedom where $\Pi_{jk}$ is the fermion polarization bubble.  In
momentum space it is given by
\cite{Ioffe}

\begin{equation}
\Pi_{jk} \approx \sqrt{J\Delta_0} \left( \delta_{jk} - \frac{q_j q_k}{q^2}
\right)  | \bbox{q} | ,
\end{equation}
{\it i.e.} it does not take the Maxwell
form which would have been proportional to $q^2$.

At higher temperatures the anisotropy term (second term) in Eq. (14) is
unimportant and the quantization
axis is disordered.  This corresponds to the spin-gap phase.  At low
temperature, the quantization axis
picks out a direction in the $x$-$y$ plane and at the same time the U(1)
symmetry corresponding to
$\bbox{a}^{(3)}$ is spontaneously broken.  This corresponds to the $d$-wave
superconductor.

Recently, quasi long-range correlations in the staggered current have been
found in the Gutzwiller projected
$d$-wave BCS wave-function \cite{Ivanov} and in the exact ground state of
small samples.\cite{Leung}  Such
current fluctuations are very natural in the SU(2) theory, and are a
consequence of fluctuations of the
quantization axis $\bbox{I}$ towards the north and south poles.  We have
suggested that these staggered
current correlations may characterize the pseudogap state, but experimental
detection of such fluctuating
currents seems to be very difficult.  Now we are ready to use this picture
to describe the vortex in the
superconducting state, and show that the staggered current fluctuations
may slow down inside the vortex core, making its detection more hopeful.  

\section{Model of the Vortex Core}

Our model of the vortex is the following.  Far away from the core $|b_1| =
|b_2|$, but $b_2 = \sqrt{x} \,z_2$ changes its phase $(\alpha + \phi/2)$ by 
$2\pi$ as we go around the
vortex, while $b_1 = \sqrt{x}\,
z_1$ does not change its phase.  The vortex contains $\frac{hc}{2e}$ flux
for the $\bbox{A}$ field and
$\frac{h}{2}$ flux for the
$\bbox {a}^{(3)}$ field.  From Eq. (15), $b_2$ sees
the sum of $\bbox{a}^{(3)}$ and $\bbox{A}$, {\it i.e.}
a unit total flux, while $b_1$ sees no net flux, so the
winding we suggested is consistent.  Note that the average phase $\alpha$
[see Eq. (11) and (12)] has a
winding of
$\pi$, as appropriate for an $\frac{hc}{2e}$ vortex.

\begin{figure}
\epsfxsize=2.0truein
\centerline{
\epsffile{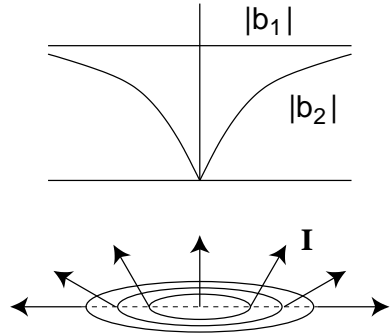}
}
\vspace{0.5cm}
\caption{Structure of the superconducting vortex.  Top: $b_1$ is constant
while $b_2$ vanishes at the
center and its phase winds by $2\pi$.  Bottom: The isospin quantization
axis points to the north pole at the
center and rotates towards the equatorial plane as one moves out radially.
The pattern is rotationally
symmetric around the $\hat{z}$ axis.}
\end{figure}

As we approach the vortex core, the amplitude of $|z_2|$ must vanish to
avoid a divergent kinetic energy
from the first term in Eq. (14).  Thus the center of the vortex core is
represented by
$\left( \tilde{h} = \sqrt{x}, 0 \right)$ and is just the staggered flux
state.  As shown in Fig. 3, the
quantization axis
$\bbox {I}$ provides a nice way to visualize this structure.  It points to
the north pole in the center of the
vortex and lies in the equator far away, but its azimuthal angle winds by
$2\pi$ as we go around the vortex.
This is sometimes referred to as the ``Meron'' configuration, or half of a
hedgehog.  It is important to recall that
$\bbox{I}$ parameterizes only the internal gauge degrees of freedom
$\theta$ and $\phi$, and the winding of
$\phi$ by $2 \pi$ has nothing to do with the winding of the overall phase
$\alpha$ by $\pi$ around the
vortex.  To visualize the winding of both $\alpha$ and $\phi$, it is
necessary to go back to the $(b_1,b_2)$ representation.

We can make a rough estimate of the vortex energy.  Assume that the core
size (size of the Meron) is
 $\ell_c$ and the size of the $\bbox{a}^{(3)}$ flux is $\ell_a$.  There are
four contributions to the energy.
The first is the energy difference between the superconducting state and
the staggered flux state.  The
main energy cost comes from the Fermi pockets.  Assuming the area of the
pockets to be $x$, we estimate an
energy cost of $\ell_c^2 x^{3/2} \sqrt{J\Delta_0}$.  On the other hand, the
Meron size cannot be smaller
than $1/\sqrt{x}$ without costing too much kinetic energy.  (In fact, the
effective action is valid only for
momenta $q \leq \sqrt{x}$ since we kept only the first quadratic term.)
The second term comes from the
electronic supercurrent and is of order $xt \ln(\lambda/\ell_c)$, where
$\lambda$ is the London penetration
 depth.  The third term comes from the supercurrent associated with the
$\bbox{a}^{(3)}$ gauge field, which
is of order $xt \ln(\ell_a/\ell_c)$, assuming $\ell_a > \ell_c$.  Finally,
the fourth contribution is from the
gauge field action, the last term in Eq. (14).  Setting $q = \ell_a^{-1}$
in Eq. (16), we estimate this
contribution to be $\ell_a^2 |a^{(3)}|^2 \sqrt{J\Delta_0}/\ell_a \approx
\sqrt{J\Delta_0}/\ell_a$.  The
important point is that unlike the U(1) case, the gauge field is not
confined to a flux tube, but can spread over
a distance
$\ell_a$.  We note that the supercurrent contributions depend
logarithmically on $\ell_a$ and
$\ell_c$, so that the main dependence comes from the first and fourth
contributions.  The staggered
flux core size
$\ell_c$ would like to be as small as possible, while the size of the gauge
flux $\ell_a$ would like to be
large.  However, our estimate of the gauge flux energy should be cut off
for $q < x$, because bosonic
contributions will enter Eq. (16).  Thus we conclude that the staggered
flux core occupies a radius of
$x^{-1/2}$ while the gauge field occupies a radius of $x^{-1}$.
The above estimate is very crude.  The main purpose is to show that a
standard $hc/2e$ vortex is possible
with a staggered flux core which does not cost too much energy as $x
\rightarrow 0$.  

If we include effects of fluctuations, the size of the staggered flux core
will very likely be bigger than the above estimate. One way to include the 
fluctuation effects is through the following consideration.
We have shown that due to the excitation of quasi-particle, the superfluid
density is reduced in the vicinity of the vortex core.\cite{LeeWen}
We have also shown that the quasi-particles carry current
$ev_F$ after including the fluctuation effects.\cite{WenLeeSC} 
In this case the superfluid density vanishes inside a
radius of $x^{-1}$, which we identify as the vortex core.\cite{LeeWen}
This argument gives a lower bound on the vortex radius, which matches the
radius $l_a$. Inside this radius the superconducting state loses phase
coherence and becomes more costly in energy. Thus our earlier estimate may
have over estimated the energy difference between the staggered flux state and
the superconducting state inside the core and
the staggered flux state may expand to 
occupies the entire core of radius $x^{-1}$ where
the superfluid density vanishes.  
The important point is the topological structure of the vortex,
which should be robust, while the details of the structure may be model
dependent. 

One important consequence of the topological structure 
is that there are two kinds of vortices, because the
isospin quantization $\bbox{I}$ can also rotate to the south pole at the
vortex core.  This just expresses the
fact that the staggered flux state is doubly degenerate, with the
staggered flux shifted by one unit cell.
In the normal state these degenerate states fluctuate between each other,
being smoothly connected via the
superconducting state.  Inside the vortex core of the superconducting state, the topological structure of the vortex forbids such smooth fluctuations,
and freezes in the staggered current pattern.  Thus the vortex
core is closely related to, but not identical to, the pseudogap state.

Since the degrees of freedom in a vortex core is finite, the two possible
staggered flux states inside the core can tunnel into each other.  If the
staggered flux core is as small as $x^{-1/2}$, the tunneling rate can be as
large as the spin gap.  However, if the staggered flux core has a size of
order $x^{-1}$ (which is more likely), the tunneling will be reduced 
exponentially. Dissipation due to quasiparticles may further suppress the
tunneling rate. 
Indeed, this problem is analogous to the 
tunneling between degenerate two level systems coupled to a Fermi sea. There 
due to the orthogonality catastrophe the tunneling rate can scale to zero and 
the state completely frozen for strong enough coupling. In general
such exponential tunneling rate is difficult to calculate, but
we are hopeful that the dynamics will slow down sufficiently for the staggered
currents to be measured experimentally.

As the magnetic field is increased, the vortex cores eventually overlap at $H
= H_{c2}$.  The staggered current states overlap and it is reasonable to
believe that the ground state should be a long range ordered staggered flux
state, especially if the staggered flux core has a size of order $x^{-1}$.
The unit cell is doubled and the ground state is a Fermi
liquid, with small Fermi pockets with area $x$.  We predicted that $H_{c2}
\sim x^2$, since the core size scales as $x^{-1}$.\cite{LeeWen}  If a high
quality underdoped sample can be made, $H_{c2}$ can be at a scale amenable to
laboratory experiment.  The Fermi pocket may be measurable by cyclotron
resonance or Shubnikov-de Haas experiments.  The cyclotron resonance has a
unique signature because the Fermi surface is close to a Dirac point so that
the Landau levels are not uniformly spaced. The doubling of the unit cell is
difficult to measure directly, because the staggered current pattern does not
couple to  charge density modulation.  It does produce a small staggered
magnetic field, which we estimate very crudely to be of order $10$
gauss.\cite{MA,Ivanov}
The possibility of detecting the staggered magnetic field by neutron
scattering and $\mu$-SR was investigated theoretically by Hsu {\it et
al}.\cite{MA} They estimated the neutron scattering intensity to be $1/70$ of
that from the ordered moments in the insulator.

\section{Experimental Probe of the Staggered Current}

If it is not possible to reach $H > H_{c2}$, the topological aspect of the
vortex offers us an opportunity to
test the staggered current picture.  It is difficult to probe the staggered
current pattern in the normal state
because of spatial and temporal fluctuations.
One of the few possible techniques is X-ray scattering
which couples to chirality fluctuations at $(\pi,\pi)$.\cite{Shastry}
However, according to our
analysis, the dynamics of the staggered current pattern slow down inside the
vortex core. Depending on the time scale, 
it may be possible to measure the
small staggered magnetic field
created by the circulating current.  
The field distribution in the vortex state is remarkably uniform, as expected
for an extreme type II superconductor.
From $\mu$-SR measurements, the field distribution has a width of roughly 5
gauss at $H = 0.5 T$.\cite{Sonier} It
should be even narrower at higher fields.
If the dynamics is slower than the $\mu$-SR scale, the field distribution
inside the vortex core is detectable. For even slower dynamics,
a more sensitive experiment is NMR.
In YBCO, the Y ion is
ideally placed to detect this current,
because it sits at the center of the plaquette.  The weak magnetic field
generated by the circulating current
will produce side bands in the Y NMR line, with a splitting independent of
$H$ but with weight proportional to
$H$.  However, there remains one complication with this proposed
experiment.  YBCO is a bi-layer
material, with Y sitting between the bi-layers.  It is likely that the
staggered pattern on the bilayers are out of
phase, in which case the magnetic field at the Y site exactly cancels.  A
way out of this difficulty is to study
the 2-4-7 structure where the two layers are asymmetric because they are
connected to different charge
reservoirs (single chain vs. double chain).  It should be possible to have
one plane of the bi-layer optimally
doped while the other plane (next to the double chain) remains underdoped.
Obviously, this proposal is quite
a challenge (but a rewarding one) for the experimentalist.

If it is possible to reach $H>H_{c2}$, NMR, $\mu$-SR, neutron, cyclotron
resonance and Shubnikov-de Haas experiments can all be performed to look for
the staggered flux state.

\section{Conclusion}

In summary, the SU(2) formulation of the $t$-$J$ model leads naturally to a
picture of the staggered flux phase above $H_{c2}$
and a stable $hc/2e$ vortex with a staggered flux core
in the superconducting state.  The basic
physical picture is that the staggered
flux state is nearly degenerate in energy to the $d$-wave
superconducting state.  The pseudogap state is
described by fluctuations between the staggered flux state and the
superconductor.  It has no long range
order, but may be characterized by short range staggered currents.  There may be short range
superconducting fluctuations as well, but these are not described
by conventional phase fluctuations alone.
As the temperature is raised above $T_c$, the fluctuations initially
resemble conventional phase fluctuations
but gradually cross over to fluctuations into the metallic staggered
flux state, all the while maintaining
the energy gap at $(0,\pi)$.  The picture may reconcile the rather
conventional $x$-$y$ model behavior
observed $30 K$ above $T_c$\cite{Corson} with the surprising persistence of
a few vortex-like excitations
up to $150 K$.\cite{Xu}

\begin{figure}
\epsfxsize=2.5truein
\centerline{
\epsffile{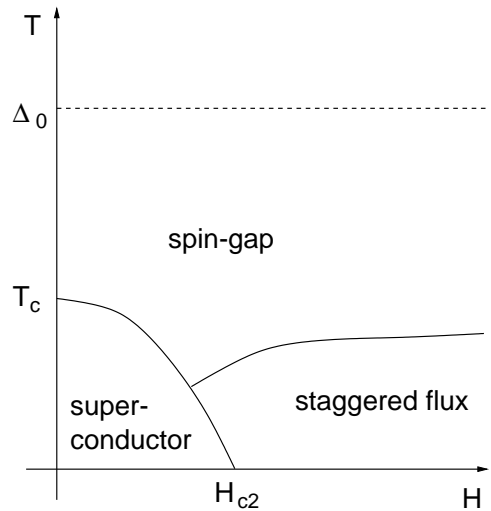}
}
\vspace{0.5cm}
\caption{Schematic phase diagram in the $(H,T)$ plane for underdoped
cuprates.  The pseudogap phase
onsets below an energy scale $\Delta_0$.  This is described by the 
spin-gap phase where the vector in Fig. 2
is disordered.  The dashed line is a cross-over temperature.  The
superconducting state appears below $T_c
\approx xt$.  Its vortex core contains ordered staggered currents.  For $H$
exceeding $H_{c2}$ the vortex
cores overlap and the staggered flux state is stabilized.}
\end{figure}

Inside the vortex core, these fluctuations are almost frozen out. The core
consists mostly the staggered flux phase
and the tunneling rate between the two kinds of vortex can be very small.
The small energy difference between the staggered flux state and the
superconductor in the limit of
small doping renders this vortex stable.  This picture suggests a $(H,T)$
phase diagram shown in Fig. 4.
Below a relatively high temperature scale (of order $\Delta_0 \approx
\frac{J}{3}$), the spin-gap phase is
formed as described above.  (This was called
the staggered-flux phase in the SU(2)
mean-field theory.\cite{WenLee})  Its onset is a
cross-over, not a phase transition.  Due to the high energy scale, this
onset is insensitive to magnetic field,
consistent with experimental findings.\cite{Gorny}  Superconductivity
onsets below a coherence temperature
$\approx xt$.  In a magnetic field, the vortex has a core of radius
$x^{-1}$. The state inside the core 
forms staggered currents on some slow time scales.  At $H_{c2} \approx x^2$,
these cores overlap, forming a truly
long range ordered staggered flux state.  This state has a doubled unit
cell and its Fermi surface consists
of small pockets of area $x$, consistent with Fermi liquid theory.  Thus
the metallic state generated by a high
magnetic field is a Fermi liquid state.  This state is connected to the
pseudogap phase by an Ising-like phase
transition.
 The long range order of the staggered flux state requires coherence
among the holes
and we expect its transition temperature to be $xt$ ({\it i.e.} comparable to the
superconducting $T_c$) as well.
This phase diagram is in contrast to a recent proposal by Chakravarty {\it
et al}.\cite{Chakravarty}, who
suggested that the onset of the pseudogap is a genuine transition.  In
their picture the staggered flux
state will extend up to the energy scale
$\Delta_0$.  The experimental test of staggered currents that we proposed should
in principle be capable of
distinguishing their proposal from ours.

We emphasize that the zero temperature ground state in the $x$-$H$ phase is
entirely conventional,
consisting of a $d$-wave superconductor, antiferromagnetic insulators, and
Fermi liquids.  At some critical
$x_c$ there is a transition between the staggered flux state with Fermi
pockets to a Fermi liquid state
with a large Fermi surface of area $(1-x)$.  The $x_c(H)$ line should
terminate at the superconducting
$H_{c2}(x)$ boundary.  Our picture of the zero temperature phase diagram
is the same as that proposed
by Chakravarty {\it et al}.  However, Chakravarty {\it et al}. asserted
that the transition between the
two Fermi liquids involves a violent change of Fermi surface topology and, by
implication, of physical properties such as
transport measurements.  In contrast, we believe that 
a line of continuous transitions with a change of the translation
symmetry is possible and in fact likely, 
in view of the smooth crossover observed at $H = 0$
above $T_c$ as a function of $x$.  The Fermi pockets are elongated and may
merge to form a single Fermi
surface in the reduced Brillouin zone in the staggered flux phase for 
$x < x_c$. The restoration of translational symmetry and
a large Fermi surface can take place
continuously by the disappearance of the Fermi surface shifted by
$(\pi,\pi)$ which lies outside the first reduced
Brillouin zone.  The scenario of a continuous evolution from small to large
Fermi surface via the ``shadow
band'' was described by one of us some time ago.\cite{PAL90}

Finally, a third alternative exists; {\it i.e.} 
the staggered flux state is never
stable (or in other words, it is destroyed by strong quantum fluctuations even
at $T=0$). In this case, something resembling the spin-gap state 
becomes the ground state in a
high magnetic field and inside the vortex core.  If true, this will be the
first example of a non-Fermi liquid ground state apart from superconductivity
in dimensions higher than one.   Our proposed phase diagram offers a very
natural route to avoid this exotic possibility.

We would like to stress that even when the vortex core is described by
the  spin-gap state, the $hc/2e$ vortex still has a small core energy which
vanishes in the $x\to 0$ limit in the mean-field theory. 
Hence the $hc/2e$ vortex is still stable. 
In fact, the $SU(2)$ vortex is the only mean-field theory at present which
gives a stable $hc/2e$ vortex with a pseudogap in the core. Whether there
exists a quasi-static staggered current inside the core is a question which is
difficult to treat theoretically, and which is best settled by experiments.

We end by making a comment on the experiments proposed by Senthil and Fisher
\cite{SF}
to test for electron fractionalization. They propose trapping a vortex in a
hole in a superconductor. When the temperature is raised above $T_c$, the
magnetic field escapes, but the $Z(2)$ vortex (vison) is trapped. Then when
the temperature is cooled down below $T_c$, the vison must capture a magnetic
flux to spontaneously form a hc/2e vortex of either sign. We would like to
point out that our model of the vortex does not exhibit the Senthil-Fisher
effect. While our vortex is also a bound state of a magnetic flux with half a
flux quantum of the gauge field $\bbox{a}^{(3)}$, the important difference is
that the gauge vortex has a finite extent and is not a flux tube. Above $T_c$,
the size of this gauge flux will expand to infinity at the same time the
size of the magnetic vortex does, {\it i.e.} the penetration depth of the
 $\bbox{a}^{(3)}$ field and the $\bbox{A}$ field both diverge in the normal
state. They allow the gauge vortex to escape the hole in the normal state.

In principle, the Senthil-Fisher effect, the electron fractionalization (or
the true spin-charge separation), and other physics of the $Z(2)$ theory can be
readily obtained from the $SU(2)$ slave boson 
theory, if one assume the $SU(2)$ gauge
symmetry is broken down to $Z(2)$ gauge symmetry (which can be achieved
by non-collinear $SU(2)$ flux through different plaquettes).\cite{WenSU2}
With this understanding, the difference between the $Z(2)$ approach and our
$SU(2)$ approach is clear. In the $Z(2)$ approach, one assumes that the $SU(2)$
gauge symmetry is broken down to $Z(2)$. While in our $SU(2)$ approach, the
$SU(2)$ is only broken down to $U(1)$ in the normal state (by a collinear 
$SU(2)$ flux). The $Z(2)$ and our $SU(2)$ approaches
correspond to different choices of mean-field states
of the same $SU(2)$ slave boson theory.

\acknowledgements{
We acknowledge helpful discussions with H. Alloul, W.
Hardy, T. Imai and N.P.
Ong concerning various experimental possibilities and limitations.  
We also thank N. Nagaosa, M. Fisher and T. Senthil for 
illuminating discussions.  P.A.L. and X.G.W.  acknowledge support by 
NSF under the MRSEC program DMR98--08941.
X.G.W also acknowledges support by NSF Grant No. DMR--97--14198.
}

\end{document}